\documentclass[
]{ceurart}

\usepackage{listings}
\usepackage{enumitem}
\usepackage{subcaption}
\usepackage{array}

\begin{document}

\copyrightyear{2026}
\copyrightclause{Copyright for this paper by its authors.
  Use permitted under Creative Commons License Attribution 4.0
  International (CC BY 4.0).}

\conference{Joint Proceedings of the ACM Intelligent User Interfaces (IUI) Workshops 2026, March 23-26, 2026, Paphos, Cyprus}

\title{Don't blame me: How Intelligent Support Affects Moral Responsibility in Human Oversight}

\author[1]{Cedric Faas}[%
orcid=0009-0000-7918-4233,
email=faas@cs.uni-saarland.de,
]
\cormark[1]

\author[2]{Richard Uth}[%
orcid=0000-0001-9697-2311,
email=richard.bergs@psychologie.uni-freiburg.de,
]

\author[1]{Sarah Sterz}[%
orcid=0000-0002-5365-2198,
email=sterz@depend.uni-saarland.de,
]

\author[2]{Markus Langer}[%
orcid=0000-0002-8165-1803,
email=markus.langer@psychologie.uni-freiburg.de,
]

\author[1]{Anna Maria Feit}[%
orcid=0000-0003-4168-6099,
email=feit@cs.uni-saarland.de,
]

\address[1]{Saarland Informatics Campus, Saarland University, Campus E 1 7, Saarbrücken, 66123, Germany}
\address[2]{Department of Psychology, University of Freiburg, Engelbergerstraße 41, Freiburg im Breisgau, 79085, Germany}

\cortext[1]{Corresponding author.}
\begin{abstract}
AI-based systems can increasingly perform work tasks autonomously.
In safety-critical tasks, human oversight of these systems is required to mitigate risks and to ensure responsibility in case something goes wrong. Since people often struggle to stay focused and perform good oversight, intelligent support systems are used to assist them, giving decision recommendations, alerting users, or restricting them from dangerous actions. However, in cases where recommendations are wrong, decision support might undermine the very reason why human oversight was employed -- genuine moral responsibility. 
The goal of our study was to investigate how a decision support system that restricted available interventions would affect overseer's perceived moral responsibility, in particular in cases where the support errs. 
In a simulated oversight experiment, participants (\textit{N}=274) monitored an autonomous drone that faced ten critical situations, choosing from six possible actions to resolve each situation. An AI system constrained participants’ choices to either six, four, two, or only one option (between-subject study). 
Results showed that participants, who were restricted to choosing from a single action, felt less morally responsible if a crash occurred. At the same time, participants' judgments about the responsibility of other stakeholders (the AI; the developer of the AI) did not change between conditions.
Our findings provide important insights for user interface design and oversight architectures: they should prevent users from attributing moral agency to AI, help them understand how moral responsibility is distributed, and, when oversight aims to prevent ethically undesirable outcomes, be designed to support the epistemic and causal conditions required for moral responsibility.
\end{abstract}


\maketitle

\section{Introduction}

AI-based systems increasingly support people at work or perform work tasks autonomously. Consider, for example, drones that can autonomously deliver packages. In such safety-critical tasks, human oversight of the system is required to mitigate risk by ensuring responsibility and preserving human judgment in decisions affecting people’s lives~\cite{jobin_global_2019, crootof_humans_2023}. If the drone is about to crash or harm others, manual control would be handed over to the overseer~\citep{lundberg_human_2021}. However, for human oversight to achieve its goals, it must be designed to be \emph{effective}, that is overseers can detect inaccurate or inadequate system outputs, malfunctions, and intervene appropriately to prevent negative effects~\cite{langer_complexities_2025}. At the same time, \citet{Sterz_2024} argue that \emph{effectiveness} also depends on the sociotechnical conditions under which oversight occurs and is closely linked to moral responsibility: to be morally responsible, overseers must satisfy both an epistemic condition (understanding why the drone crashes and how it could have been prevented) and a causal condition (having caused or contributed to the crash)~\cite{talbert_moral_2024,noorman_computing_2023}.

A potential approach to enhance overseers' epistemic access, decision accuracy, and overall safety \citep{lundberg_human_2021, URAIKUL2007115} is to provide additional information or specific decision recommendations~\citep{bansal_does_2021, bucinca_proxy_2020, green_principles_2019, poursabzi-sangdeh_manipulating_2021, zhang_effect_2020}. These enhancements may indirectly support responsibility by helping the overseer better understand their decision situation and prevent harm more effectively. However, in high-stakes scenarios, improving cognitive factors might not be enough. If the overseer gets bored or distracted while monitoring the drone, they will struggle to maintain situational awareness~\citep{Parasuraman_2010}, and there might be strong reasons to ensure safety by restricting the overseer's freedom of choice in addition to increasing their epistemic access. In the drone example, safety could be increased by restricting the overseer from selecting particularly dangerous control options. However, such restrictions also reduce human control, diminish the overseer's moral responsibility by weakening the causal connection between their actions and potential crashes, and, in extreme cases, relegate them to the role of a scapegoat if something goes wrong~\cite{Green_2022, elish_moral_2019}. Empirical research raises doubts about whether human oversight achieves the desired performance~\cite{Green_2022, afroogh_task-driven_2025, Vaccaro_2024}, which raises normative concerns about whether humans can genuinely fulfill their assigned responsibility under certain constraints. This leaves us with a dilemma: restricting the overseer’s freedom of action can improve safety and performance, yet this decision support may undermine the very reason why human oversight was employed -- genuine moral responsibility.

The goal of this paper is to shed light on this dilemma and to investigate how restrictive decision support that ensures safety affects the overseer's judgments of moral responsibility. Therefore, we present results from a between-participant study where participants performed an oversight task, monitoring an autonomous delivery drone. During the study, the drone faced different critical situations that required the overseer to choose an action to resolve the critical situation (e.g., bad weather conditions) safely. Participants received decision support from a (simulated and perfectly accurate) AI system, either by showing \emph{Six} possible actions to resolve the situation or by making only \emph{Four} or \emph{Two}, or \emph{One} of these actions selectable while the other options were grayed out (i.e., not selectable, due to the AI categorizing them as unsafe). This restriction increased the overall safety of the process, since the correct action was always recommended. We hypothesized that restricting the number of \textit{Selectable Actions} would affect participants' responsibility judgments.

\section{Method}
\subsection{Participants}
A-priori power analysis (G*Power; \citep{faul_gpower_2007}) determined a required sample size of \textit{N}~=~271 to achieve a power of 1-$\beta$ = .95 at $\alpha$ = .05, assuming a medium effect size (\textit{f}\,=\,.25). We recruited 280 participants via Prolific (151 female, 129 male; age: 18-71, \textit{M}~=~33.25, \textit{SD}~=~9.81). After data cleaning in accordance with our preregistration, the final sample consisted of \textit{N}~=~274. Our study was approved by the first author's institution's ethical review board and preregistered via the \href{https://osf.io/r2jhw/?view_only=2d18c3ea8c9849eab6b4004e93ce0909}{Open Science Foundation}, where we also published the collected data. 

\begin{figure}[h]
    \centering
    \includegraphics[width=1\linewidth]{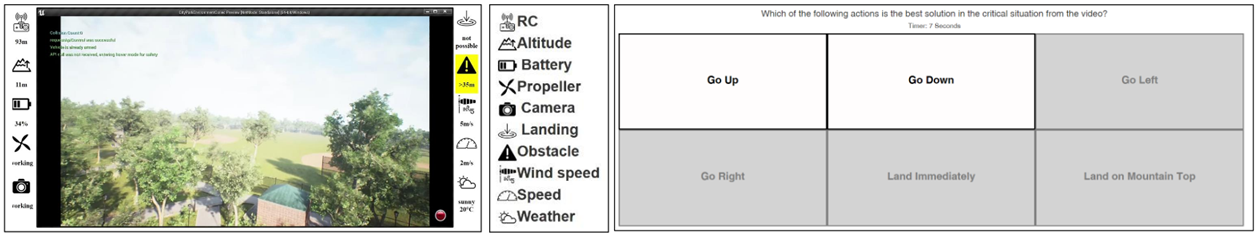}
    \caption{Screenshots of the drone monitoring interface. Left: Screenshot of the demo task, where participants saw a video recorded by the drone and information about the current status. After thirty seconds, the drone entered a critical situation, indicated by an auditory signal and highlighted icons for critical values. Middle: Legend of all icons used in the interface and introduced during the training phase; it was shown to the participants during the entire video. Right: After ten seconds, participants saw six possible actions, and depending on the experimental condition, some of them were grayed out by an AI decision support indicating they were not available (here: \emph{Two} \textit{Selectable Actions} condition).}
    \label{fig: task}
\end{figure}

\subsection{Procedure}
At the start, participants received instructions on the task, sensor values, critical situations, and possible actions, followed by a trial task. In the main task, they watched ten 40-second drone flight videos. When it came to a critical situation, an auditory signal indicated a handover to the human overseer. After ten seconds, the video stopped, and six actions were displayed, with some grayed out depending on the condition. Participants had up to seven seconds to select an action or reject all. This task provided a realistic yet relatable decision-making scenario under time constraints while maintaining experimental control to test our hypotheses. The interface and study setup ensured no prior drone expertise was needed. See \autoref{fig: task} for screenshots of the interface and selectable actions.
After the last trial, participants filled out a questionnaire with Likert-scale items assessing their judgments of moral responsibility. In the final trial, all actions, even though some actions seemed plausible, would lead to a crash. This was done to ensure that participants experienced at least one undesired outcome in all conditions. See \citet{faas_give_2024}, for more details on the procedure and findings on additional questions.

\subsection{Measures}
We used self-report items to measure perceived \textit{moral responsibility of the oversight person}, \textit{moral responsibility of the system}, \textit{moral responsibility of the developer of the system},  \textit{causality}, and \textit{knowledge} on a 7-point Likert scale (1 - strongly disagree to 7 - strongly agree). All items and further exploratory variables are presented in \autoref{app:measures} along with references to the literature from which we sourced these items.

We additionally captured performance-related variables: \textit{decision accuracy} and \textit{decision time}. Decision accuracy was calculated for each participant as the percentage of selected actions that led to a successful outcome (i.e., drone not crashing) in the first nine rounds. Decision time was measured as the time from when participants first saw the available actions until they chose a specific action.

\section{Results}

\begin{figure}[h!]
\centering
\begin{subfigure}{0.25\linewidth}
\includegraphics[width=\linewidth]{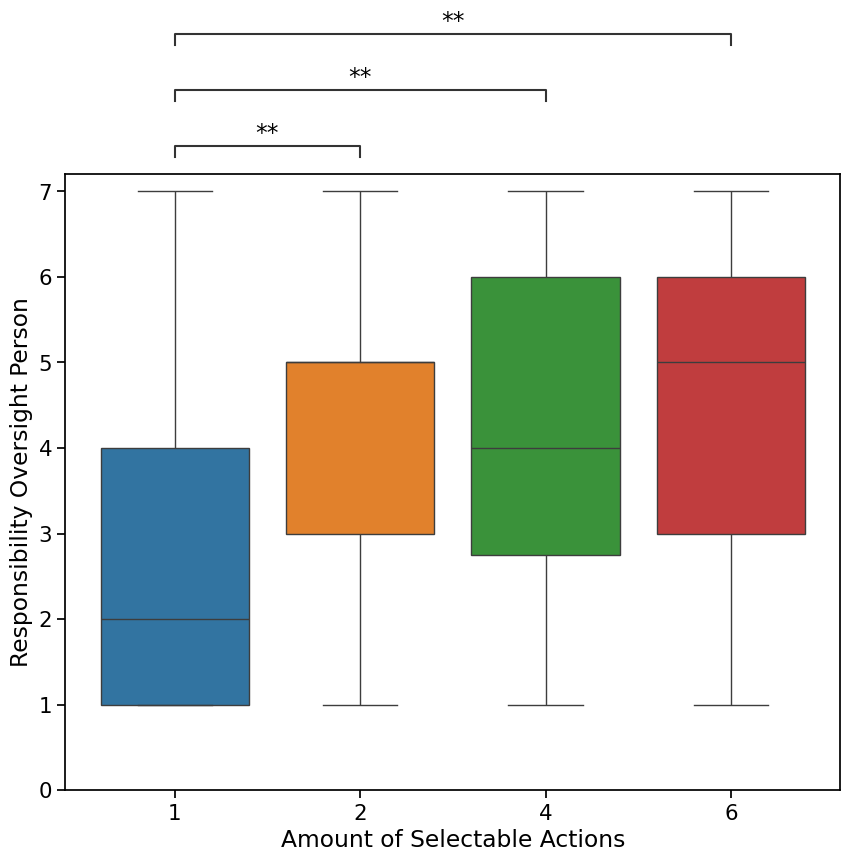}
\caption{Moral Responsibility}
\label{fig:boxplot_Responsibility_OP}
\end{subfigure}
\begin{subfigure}{0.25\linewidth}
\includegraphics[width=\linewidth]{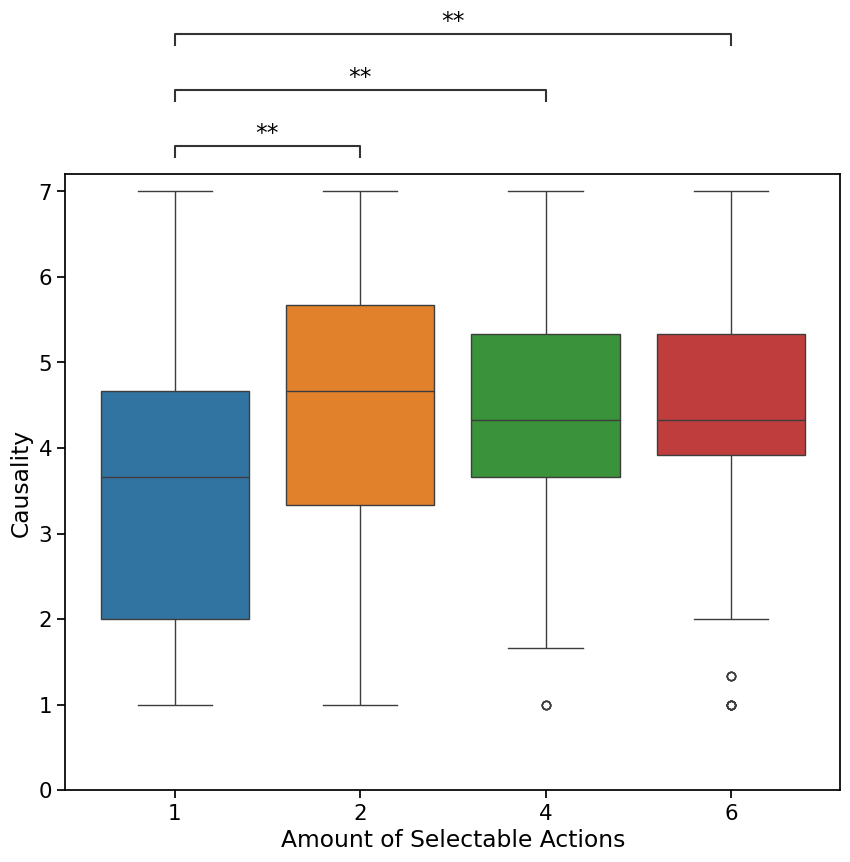}
\caption{Causality}
\label{fig:boxplot_Causality}
\end{subfigure}
\begin{subfigure}{0.25\linewidth}
\includegraphics[width=\linewidth]{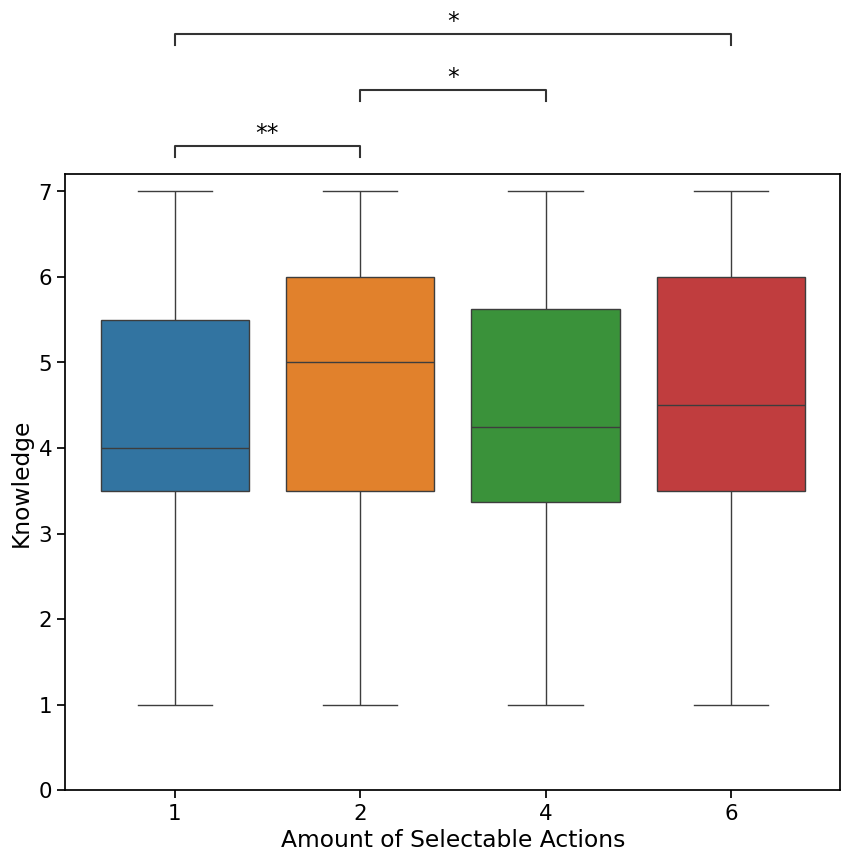}
\caption{Knowledge}
\label{fig:boxplot_Knowledge}
\end{subfigure}\\
\footnotesize Note: *p<=0.05, **p<=0.01 
\caption{Differences in participants' own responsibility, causality, and knowledge judgments on a 7-point Likert Scale between the experimental conditions}
\label{fig:boxplotCausalityKnowledge}
\end{figure}

We report the overall effects of \textit{Selectable Actions} on the participants' judgements about \textit{moral responsibility of the oversight person}, \textit{moral responsibility of the system}, \textit{moral responsibility of the developer of the system}, \textit{causality} and \textit{knowledge}. To test for significant differences, we chose a Kruskal-Wallis test combined with the Dunn test with Bonferroni correction for the post-hoc comparisons of all experimental conditions. Figures with boxplots can be found in \autoref{fig:boxplotCausalityKnowledge} and \autoref{app:boxplots}. 

\emph{Selectable Actions} significantly affected participants' perceived \textit{moral responsibility of the oversight person} (H~=~98.87, p~<~.001) and \textit{moral responsibility of the system} (H~=~11.35, p~<~.01). 
With \emph{One} \textit{Selectable Action}, participants felt less responsible than having \emph{Two} (p~=~.01), \emph{Four} (p~=~.01), or \emph{Six} (p~=~.01 ) \emph{Selectable Actions} (\autoref{fig:boxplot_Responsibility_OP}).
Participants' judgment about the system's responsibility varied more when having \emph{One} \textit{Selectable Action} compared to having \emph{Six} (p~<~.02) \emph{Selectable Actions}.
In contrast, there were no significant differences for \textit{moral responsibility of the developer of the system} across the conditions. 
Surprisingly, for all stakeholders there was no significant variance in the responsibility judgments between the \emph{Two}, \emph{Four}, or \emph{Six} \emph{Selectable Actions} conditions.
Furthermore, \emph{Selectable Actions} significantly affected the participants perceived \textit{causality} (\autoref{fig:boxplot_Causality}) (H~=~58.58, p~<~.001) and \textit{knowledge} (\autoref{fig:boxplot_Knowledge}) (H~=~18.56, p~<~.001). 

The number of \emph{Selectable Actions} also significantly affected the decision accuracy (H~=~128.88, p~<~.001). Pairwise post-hoc comparisons showed, as expected, participants having \emph{One} \textit{Selectable Action} to make better decisions than participants having \emph{Two} (p~<~.001), \emph{Four} (p~<~.001), or \emph{Six} (p~<~.001)  \emph{Selectable Actions}. Surprisingly, no significant differences were found when comparing conditions with more than \emph{One} \emph{Selectable Actions}, thus decision accuracy did not decrease with an increasing number of selectable actions.
Notably in all conditions, a small number of decisions (\emph{One}: 22, \emph{Two}: 11, \emph{Four}: 6, \emph{Six}: 11, out of 700 per Condition) reflected that participants deliberately \emph{rejected} all actions.
We also found a significant difference in the \textit{decision time} between the experimental conditions (H~=~28.79, p~<~.001). Participants who had \emph{One} \textit{Selectable Action} decided faster than those with \emph{Two} (p~<~.001), \emph{Four} (p~<~.001), or \emph{Six} (p~<~.001) \emph{Selectable Actions}. No significant differences were found when comparing conditions with more than \emph{One} \emph{Selectable Actions}.

After each decision, the participants were asked three open-ended questions about the reasoning behind their decisions (see \autoref{app:measures}). Their answers showed that the participants stayed engaged during the entire study and that they did not select actions at random. The answers also showed that participants understood the criteria to base their decisions on and that they were able to provide reasons for their decisions, e.g., "The camera broke, I couldn't continue the flight path" [Four selectable actions; Trial  6], even though the decision accuracy results showed that the task was challenging for them. 

\section{Discussion \& Conclusion}
We see three main limitations of our study. First, participants only experienced a simulated drone oversight task without any real-world consequences. Nevertheless, we tried to simulate a realistic situation by showing participants videos of drone flights and by simulating time pressure. 
Second, participants presumably had limited experience in flying drones and were not experts in human oversight tasks. This lack of expertise might have affected their overall perceived responsibility. Still, we believe that the current study included a task where our participants were able to immerse themselves and quickly judge their responsibilities. 
Lastly, we only measured the perceived responsibility of the participants, the system, and the developer. There might be further stakeholders that can be held morally responsible for undesirable outcomes (e.g., the deployer of the system or the overseer's organization).

Overall, our results highlight a central trade-off: restricting choices can boost safety and accuracy \citep{Grote_2020, Lai_2023, eisbach_optimizing_2023}, but risks undermining the overseer's moral responsibility. 
Although potentially conflicting with design guidelines \citep{amershi_guidelines_2019, li_assessing_2023,jobin_global_2019}, this underscores that removing ineffective options can be a valid system design option. 
While limiting choice to approving or ignoring system recommendations may only be appropriate for specific oversight contexts, it might align with emerging regulatory requirements, such as those in the EU AI Act, that mandate human involvement for high-risk AI systems~\citep{Sterz_2024, Green_2022}.

An important challenge is to balance removing ineffective choices and preserving a meaningful decision. Our findings suggest that restricting ineffective options does not necessarily diminish perceived responsibility, as long as there is some meaningful choice left. This is promising for safety-critical domains where minimizing errors is essential~\citep{Grote_2020, lundberg_human_2021}. Thus, designers may limit user choices to improve performance, as long as users retain a meaningful decision. However, this conclusion comes with ethical risks: designers may be tempted to leave users some choice by providing pseudo-choices. As \citet{Sterz_2024} argue, oversight based on pseudo-choices is neither effective nor morally responsible and should not satisfy regulatory requirements, such as the EU AI Act. Our findings support this view: participants felt morally responsible only when they had a meaningful choice, and deliberately choosing inaction was rare (44 of 2,800 decisions), likely because it constituted a pseudo-choice that could not resolve the situation. Moreover, our simulated AI always removed the most unsuitable options. If an AI were to restrict options to one suitable and one unsuitable action, oversight would become ineffective, making it unreasonable to hold the human morally responsible. On the other hand, in our study, performance improvements were only detected when no meaningful choice was left. This raises the question of whether human oversight should be required when safety-oriented interaction designs undermine overseers’ responsibility without meaningfully benefiting from human involvement. This concern aligns with emerging criticism of legal requirements for human oversight~\citep{Green_2022} and underscores the need to define what constitutes effective oversight in ways that connect to the moral responsibility~\citep{Sterz_2024}.

Responsibility is typically assigned to enable praise or blame. In the case of drone oversight, the overseer may be praised for mitigating harm or blamed for causing it. Still, other stakeholders may also bear moral responsibility.
Our findings indicate that participants who had a choice attributed a shared responsibility between themselves, the AI system, and its developer. Across all conditions, participants held the AI and Developers equally responsible, but those who did not have a choice felt less responsible. This suggests that participants treated the AI as a moral agent, which conflicts with most theories of moral agency. Although increasing autonomy and capability may encourage people to view AI systems as moral agents, dominant philosophical accounts hold that moral agency is exclusive to humans who possess specific powers and capabilities, which are typically attributed to adult humans, but not to very young children, non-human animals, or computer systems~\cite{talbert_moral_2024, noorman_computing_2023}. If oversight personnel attributed moral agency to the overseen AI, they might falsely assume moral considerations in the AI's recommendations. This perception of the AI's moral agency could influence their decision-making, which is already heavily influenced by AI behavior in morally challenging situations~\cite{salatino_influence_2025}. Therefore, oversight design should align with philosophical accounts of moral agency and responsibility~\cite{faas_design_2025}.

Regardless of the debates about who qualifies as a moral agent, moral responsibility typically requires both a causal and an epistemic condition~\cite{talbert_moral_2024, noorman_computing_2023}. An agent can be held responsible for an undesirable outcome only if their actions causally lead to it and if they possess sufficient understanding of the decision context. This framework aligns with our findings: participants without a meaningful choice felt less knowledgeable and perceived a weaker causal link to the drone crash compared to those with a choice. This insight can have different implications depending on the goal of human oversight or the reasons why it is employed. 
If the main goal of Human Oversight is to prevent ethically undesirable outcomes, interface design should prioritize enhancing the overseer’s epistemic access and ensuring that they can establish a sufficient causal connection to the outcomes of the system. Overseers need support in understanding the decision situation, the AI's capabilities, and its strengths and limitations~\cite{apple_human_nodate,google_google_2019,amershi_guidelines_2019,faas_design_2025}. Although decision support can increase epistemic access~\citep{Das_2024, Parasuraman_2010, herm_impact_2023}, designers should ensure meaningful choice and the possibility to prevent undesirable outcomes~\cite{Sterz_2024}.
When Human Oversight is used primarily for performance improvements, limited epistemic access and causal contribution may be less critical. Even then, however, clearly communicating responsibilities remains essential, both so overseers understand the boundaries of their own responsibility if something goes wrong and to prevent them from being positioned as scapegoats~\cite{faas_design_2025, Green_2022, elish_moral_2019}.

\begin{acknowledgments}
  This work was funded by DFG grant 389792660 as part of TRR~248 -- CPEC, see \url{https://perspicuous-computing.science}. We thank Emilia Ellsiepen for all the support regarding the study design and execution.
\end{acknowledgments}

\section*{Declaration on Generative AI}
 During the preparation of this work, the authors used Chat-GPT and Grammarly in order to: Grammar and spelling check. After using these tools, the authors reviewed and edited the content as needed and take full responsibility for the publication’s content.

\bibliography{bibliography}

\appendix
\newpage
\section{Measures}
\label{app:measures}
\subsection{Questionnaire Items}
\begin{table*}[h]
    \centering
\caption{Questionnaire items and measured concepts}
\label{tab:questionnaire_items}

\resizebox{\textwidth}{!}{
    \begin{tabular}{>{\raggedright\arraybackslash}p{0.3\linewidth} >{\raggedright\arraybackslash}p{0.6\linewidth} c} \toprule
        Concept & Question(s)&Ref\\ \midrule
        Moral Responsibility of the Oversight Person & 'I was responsible for the crash of the drone.' & \cite{hinds_whose_2004}\\ 
        Moral Responsibility of the System & 'The system was responsible for the crash of the drone.' & \cite{hinds_whose_2004}\\ 
        Moral Responsibility of the Developer of the System & 'The developer of the system was responsible for the crash of the drone.' & \cite{hinds_whose_2004}\\ \midrule
        \multicolumn{3}{l} {Additional exploratory questionnaire items and measured concepts}\\ \midrule
        Perceived Autonomy & 'I am so restricted by the available actions that I can hardly make my own decisions.' \textit{(inversely coded)} &  \cite{van_dick_job_2001}\\
        &  'I can make decisions about how to prevent crashes independently.'&\\ 
        Perceived Meaningfulness & 'The work I do is important.'  &\cite{spreitzer_empirical_1995}\\
         & 'The work I do is meaningful.'& \\ 
        Satisfaction &'I feel fulfilled performing the task.' &\cite{parker_measurement_2011}\\
        Motivation &'I have fun performing the task.'  &\cite{gagne_motivation_2010}\\
        Causality& 'I caused the drone to crash.' & \cite{gailey_attribution_2008}\\
        &'I could have avoided crashing the drone.' & \\
        &'I could have prevented the drone from crashing.' & \\
        Knowledge&'I could have foreseen the crash of the drone.'  &\cite{gailey_attribution_2008}\\
        &'I understand why the drone crashed.' & \\
        Intention&'I had the intention of crashing the drone.' &\cite{gailey_attribution_2008}\\
        Control&'I felt like I was in control over the crash of the drone.'   & \cite{nolan_threat_2016}\\
        Trust&'I can trust the system.'  & \cite{korber_theoretical_2019}\\\
        &'I can rely on the available options.' &\\
        Disengagement&'I feel that I have become disconnected from my work task. ' &\cite{demerouti2008oldenburg}\\
        Exhaustion&'I feel tired performing the task. ' &\cite{demerouti2008oldenburg}\\
        Stress&'I am currently stressed by the task.'   & \cite{motowidlo_occupational_1986} \\
        Blame&'To what extent do you think you are to blame for the drone crash?   & \cite{ames_intentional_2013} \\
        Punishment&'To what extent should you be punished for the crash of the drone?'   & \cite{yao_influence_2021}\\
        Task Load (repeated subset) &'The task was mentally demanding.'    &\cite{hart_development_1988}\\
        &'The pace of the task was hurried or rushed.' & \\
        Performance&'I was successful in accomplishing the task.'   & \cite{hart_development_1988}\\
        Physical Demand&'The task was physically demanding.' & \\
        Effort&'I had to work hard to accomplish my level of performance.'  & \\
        Frustration&'I was irritated, stressed, and annoyed during the task.'   &\\\bottomrule
    \end{tabular}}
\end{table*}

\subsection{Open Questions}
\label{app:openQuestions}
Open ended questions, the participants were asked about their reasoning after each decision:
\begin{enumerate}[label={Q\arabic*.}]
    \item Which information did you consider when selecting this option and why?
    \item What about the other options? Are there any reasons why they are not as suitable?
    \item Would you select any of the other actions and why?
\end{enumerate}

\newpage
\section{Visualization of Findings}
\label{app:boxplots}

\begin{figure}[h]
\centering
\begin{subfigure}{0.33\linewidth}
\includegraphics[width=\linewidth]{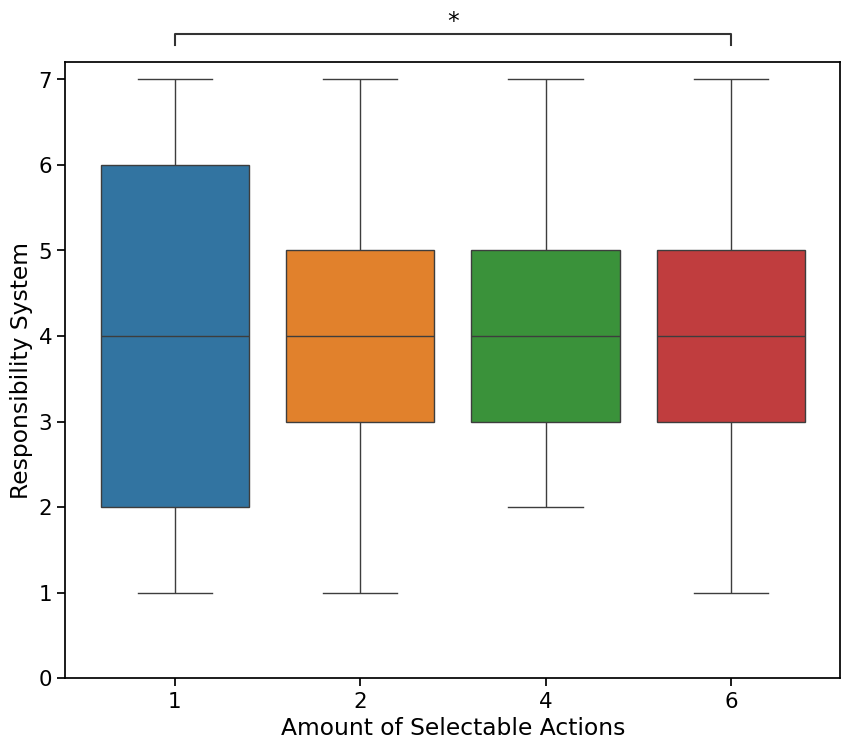}
\caption{Moral Responsibility System}
\label{fig:boxplot_Responsibility_S}
\end{subfigure}
\begin{subfigure}{0.33\linewidth}
\includegraphics[width=\linewidth]{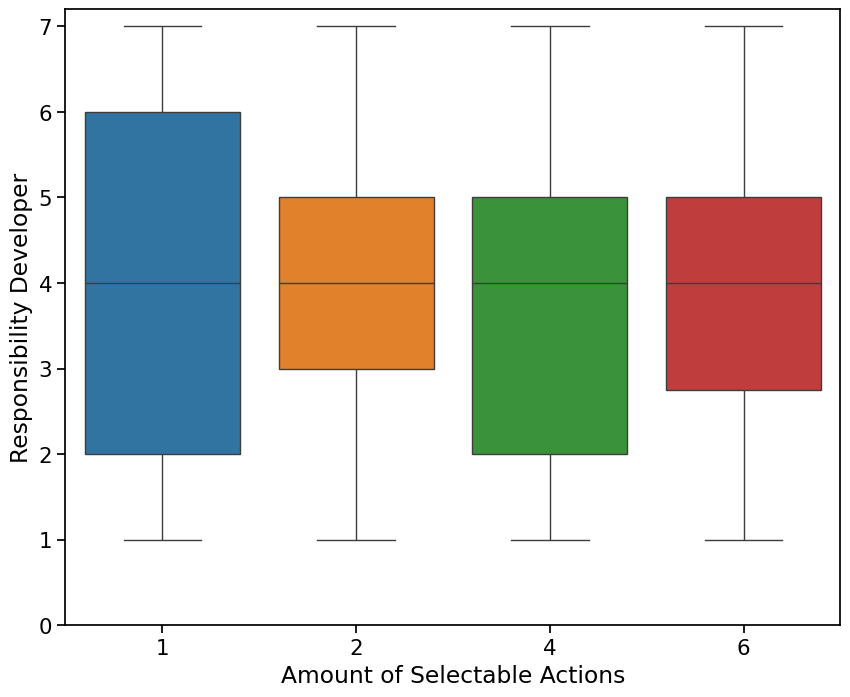}
\caption{Moral Responsibility Developer}
\label{fig:boxplot_Responsibility_D}
\end{subfigure}\\
\footnotesize Note: *p<=0.05, **p<=0.01
\caption{Differences in responsibility judgments on a 7-point Likert Scale between the experimental conditions} 
\label{fig:boxplotAppendix}
\end{figure}

\begin{figure}[h]
\centering
\begin{subfigure}{0.25\linewidth}
\includegraphics[width=\linewidth]{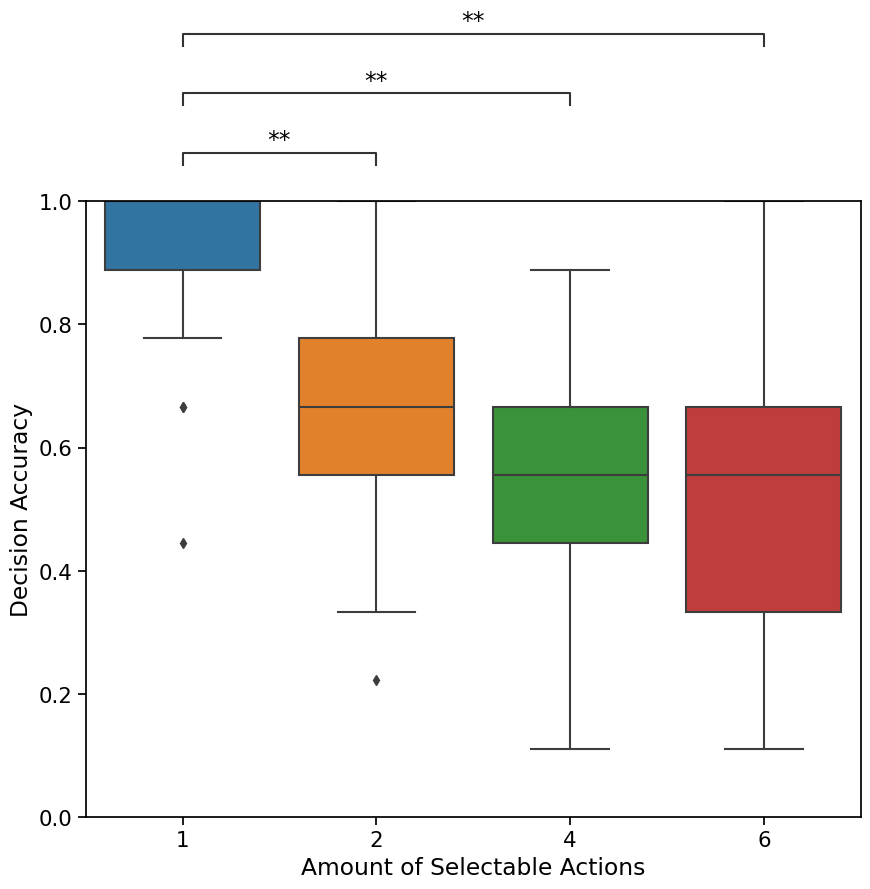}
\caption{Decision Accuracy}
\label{fig:boxplot_DecisionAccuracy_Conditions}
\end{subfigure}
\begin{subfigure}{0.25\linewidth}
\includegraphics[width=\linewidth]{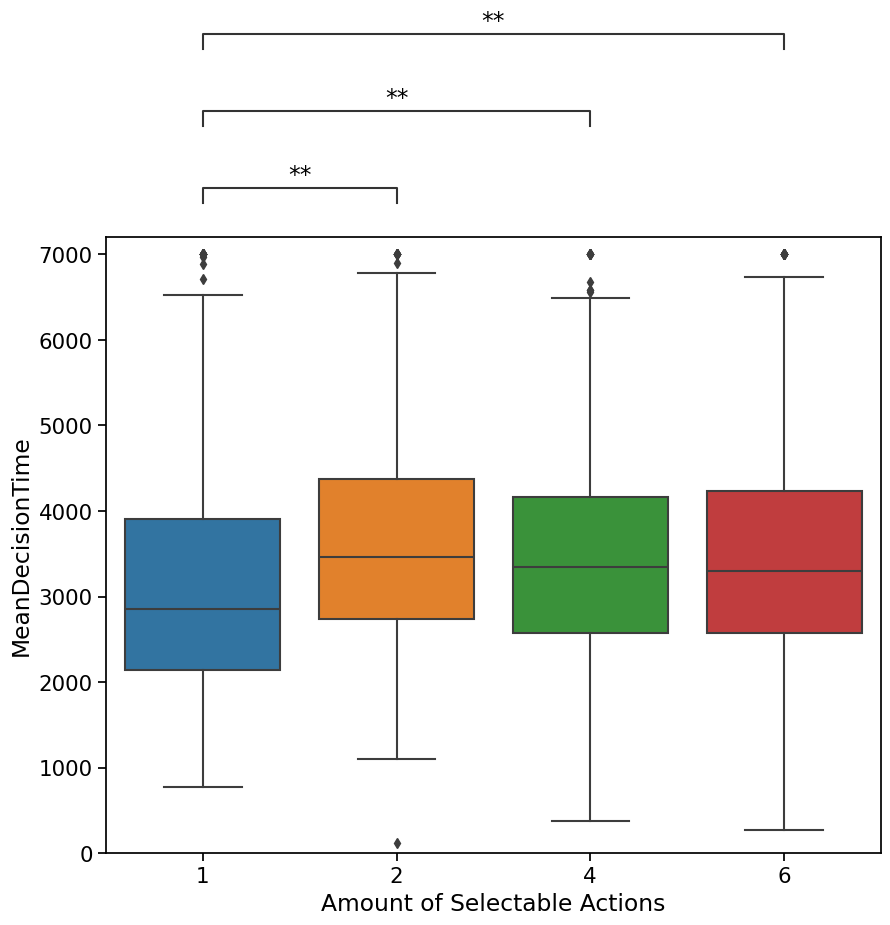}
\caption{Decision Time in ms}
\label{fig:boxplot_DecisionTime_Conditions}
\end{subfigure}\\
\footnotesize Note: *p<=0.05, **p<=0.01 
\caption{Differences in decision accuracy in percentage and decision time in milliseconds between the experimental conditions}
\label{fig:boxplotDecisionTimeAccuracy}
\end{figure}

\end{document}